\documentclass[prl,superscriptaddress,twocolumn,longbibliography]{revtex4-1}
\usepackage{amsmath,amssymb}
\usepackage{subfigure,bbm,times}
\usepackage[T1]{fontenc}
\usepackage{braket}
\usepackage{graphicx}
\usepackage{float}
\usepackage{verbatim}
\usepackage{url}

\usepackage{array,amsfonts}
\usepackage{dcolumn,tabularx,booktabs}
\usepackage{bm}
\usepackage{color}
\usepackage[colorlinks,citecolor=blue]{hyperref}
\usepackage{bbold}

\begin{document}
	\hyphenpenalty=5000
	\tolerance=1000

\title{High-fidelity realisation of CNOT gate in Majorana-based optical platform}

\author{Jia-Kun Li}
\affiliation{Laboratory of Quantum Information, University of Science and Technology of China, Hefei 230026, China}
\affiliation{Anhui Province Key Laboratory of Quantum Network, University of Science and Technology of China, Hefei 230026, China}
\affiliation{CAS Center for Excellence in Quantum Information and Quantum Physics, University of Science and Technology of China, Hefei 230026, China}

\author{Kai Sun}
\email{ksun678@ustc.edu.cn}
\affiliation{Laboratory of Quantum Information, University of Science and Technology of China, Hefei 230026, China}
\affiliation{Anhui Province Key Laboratory of Quantum Network, University of Science and Technology of China, Hefei 230026, China}
\affiliation{CAS Center for Excellence in Quantum Information and Quantum Physics, University of Science and Technology of China, Hefei 230026, China}

\author{Ze-Yan Hao}
\affiliation{Laboratory of Quantum Information, University of Science and Technology of China, Hefei 230026, China}
\affiliation{Anhui Province Key Laboratory of Quantum Network, University of Science and Technology of China, Hefei 230026, China}
\affiliation{CAS Center for Excellence in Quantum Information and Quantum Physics, University of Science and Technology of China, Hefei 230026, China}

\author{Jia-He Liang}
\affiliation{Laboratory of Quantum Information, University of Science and Technology of China, Hefei 230026, China}
\affiliation{Anhui Province Key Laboratory of Quantum Network, University of Science and Technology of China, Hefei 230026, China}
\affiliation{CAS Center for Excellence in Quantum Information and Quantum Physics, University of Science and Technology of China, Hefei 230026, China}

\author{Jiannis K. Pachos}
\email{j.k.pachos@leeds.ac.uk}
\affiliation{School of Physics and Astronomy, University of Leeds, Leeds LS2 9JT, United Kingdom}

\author{Lucy Byles}
\affiliation{School of Physics and Astronomy, University of Leeds, Leeds LS2 9JT, United Kingdom}

\author{Jin-Shi Xu}
\email{jsxu@ustc.edu.cn}
\affiliation{Laboratory of Quantum Information, University of Science and Technology of China, Hefei 230026, China}
\affiliation{Anhui Province Key Laboratory of Quantum Network, University of Science and Technology of China, Hefei 230026, China}
\affiliation{CAS Center for Excellence in Quantum Information and Quantum Physics, University of Science and Technology of China, Hefei 230026, China}
\affiliation{Hefei National Laboratory, University of Science and Technology of China, Hefei 230088, China}

\author{Yong-Jian Han}
\email{smhan@ustc.edu.cn}
\affiliation{Laboratory of Quantum Information, University of Science and Technology of China, Hefei 230026, China}
\affiliation{Anhui Province Key Laboratory of Quantum Network, University of Science and Technology of China, Hefei 230026, China}
\affiliation{CAS Center for Excellence in Quantum Information and Quantum Physics, University of Science and Technology of China, Hefei 230026, China}
\affiliation{Hefei National Laboratory, University of Science and Technology of China, Hefei 230088, China}

\author{Chuan-Feng Li}
\email{cfli@ustc.edu.cn}
\affiliation{Laboratory of Quantum Information, University of Science and Technology of China, Hefei 230026, China}
\affiliation{Anhui Province Key Laboratory of Quantum Network, University of Science and Technology of China, Hefei 230026, China}
\affiliation{CAS Center for Excellence in Quantum Information and Quantum Physics, University of Science and Technology of China, Hefei 230026, China}
\affiliation{Hefei National Laboratory, University of Science and Technology of China, Hefei 230088, China}

\author{Guang-Can Guo}
\affiliation{Laboratory of Quantum Information, University of Science and Technology of China, Hefei 230026, China}
\affiliation{Anhui Province Key Laboratory of Quantum Network, University of Science and Technology of China, Hefei 230026, China}
\affiliation{CAS Center for Excellence in Quantum Information and Quantum Physics, University of Science and Technology of China, Hefei 230026, China}
\affiliation{Hefei National Laboratory, University of Science and Technology of China, Hefei 230088, China}

\begin{abstract}
We present the experimental realisation of a robust CNOT quantum gate using Majorana zero modes simulated on a photonic platform. Three Kitaev chains supporting Majorana zero modes at their endpoints are used to encode two logical qubits, and both intra-chain and inter-chain braiding operations are performed to implement the CNOT gate. While the topological encoding of quantum information in Majorana fermions does not offer full topological protection in our non-interacting photonic setting, it nevertheless exhibits a natural resilience to the dominant noise and decoherence effects present in the experiment. Consequently, the fidelity of the CNOT gate is significantly enhanced, surpassing 0.992 and addressing a key limitation in the path toward scalable quantum computation. These results represent a major advancement in topological quantum computing with Majorana fermions and underscore the potential of photonic platforms for realising high-fidelity quantum gates.
\end{abstract}

\maketitle

\noindent
\textbf{Introduction}

Quantum computation provides a paradigm that clearly differs from and is conceivably far more powerful than established classical computation \cite{Feynman1982}, due to its quantum characteristics such as superposition and entanglement \cite{RevModPhys.85.1103}. The cornerstone of implementing  quantum computation is quantum ``gate array'' \cite{Deutsch1989}, which is composed of a set of elementary quantum gates \cite{PhysRevA.52.3457} . Among these gates, the quantum controlled-NOT (CNOT) gate stands out as a fundamental element in quantum computation. It has been proven that universal quantum computation can be achieved by combining a series of simple single-qubit gates with the CNOT gates, in which all the unitary operations on arbitrarily n-qubit can be constructed using these basic gates \cite{Barenco1995}. The frequently utilized entangled Bell states for quantum information processing can also be generated through the application of the CNOT gates \cite{PhysRevLett.126.130501,Zhang2023,Qiang2018}. A significant amount of study, both theoretically and experimentally, has been conducted surrounding CNOT gates in varied physical systems, such as nuclear magnetic resonance \cite{PhysRevLett.110.190501,PhysRevB.93.035306,PhysRevA.97.022311}, nitrogen-vacancy (NV) centers \cite{PhysRevA.90.052310,Du24,Rong2015}, trapped ions \cite{PhysRevLett.74.4091,PhysRevLett.81.3631,Schmidt-Kaler2003},  photonic integrated circuits \cite{Shadbolt2012,Zeuner2018,Zhang2019} and linear optics \cite{Knill2001,O'Brien2003,PhysRevLett.95.210506}. The central objective of these investigations is to realise the CNOT gate with high fidelity, as it typically constitutes the primary source of decoherence and thus poses a major bottleneck to the practical scalability of quantum computing.

Topological quantum computing has attracted significant attention for its potential to overcome the limitations of conventional quantum computation, particularly the susceptibility to errors and decoherence \cite{KITAEV20032,PhysRevLett.94.166802}. By utilising Majorana zero modes (MZMs), which exhibit non-Abelian statistics, quantum information can be stored in a topologically protected manner \cite{RevModPhys.87.137,PhysRevLett.116.257003,RevModPhys.83.1057}. This protection makes MZMs ideal candidates for fault-tolerant quantum computing \cite{PhysRevLett.98.190504,PhysRevA.57.127,PhysRevA.63.042307}, where local perturbations have minimal impact on encoded information. While universal quantum computation cannot be realized by simply braiding of MZMs \cite{RevModPhys.80.1083}, the introduction of an additional non-Clifford gate ($\pi$/8-phase gate) \cite{PhysRevLett.104.180505} can resolve this issue. Despite the $\pi$/8-phase gate's vulnerability to local noise, the application of ``magic state distillation'' \cite{PhysRevA.71.022316} offers a pathway to its error correction. Meanwhile, a universal set of MZMs based single-qubits has been demonstrated experimentally in linear optics system \cite{Xu2016,Xu2018}, leaving only the pivotal two-qubit CNOT gates as the component for realizing universal topological quantum computation within the MZMs-based optical platform. 

\begin{figure}[htbp]
 	\centering
 	\includegraphics[width=1\columnwidth]{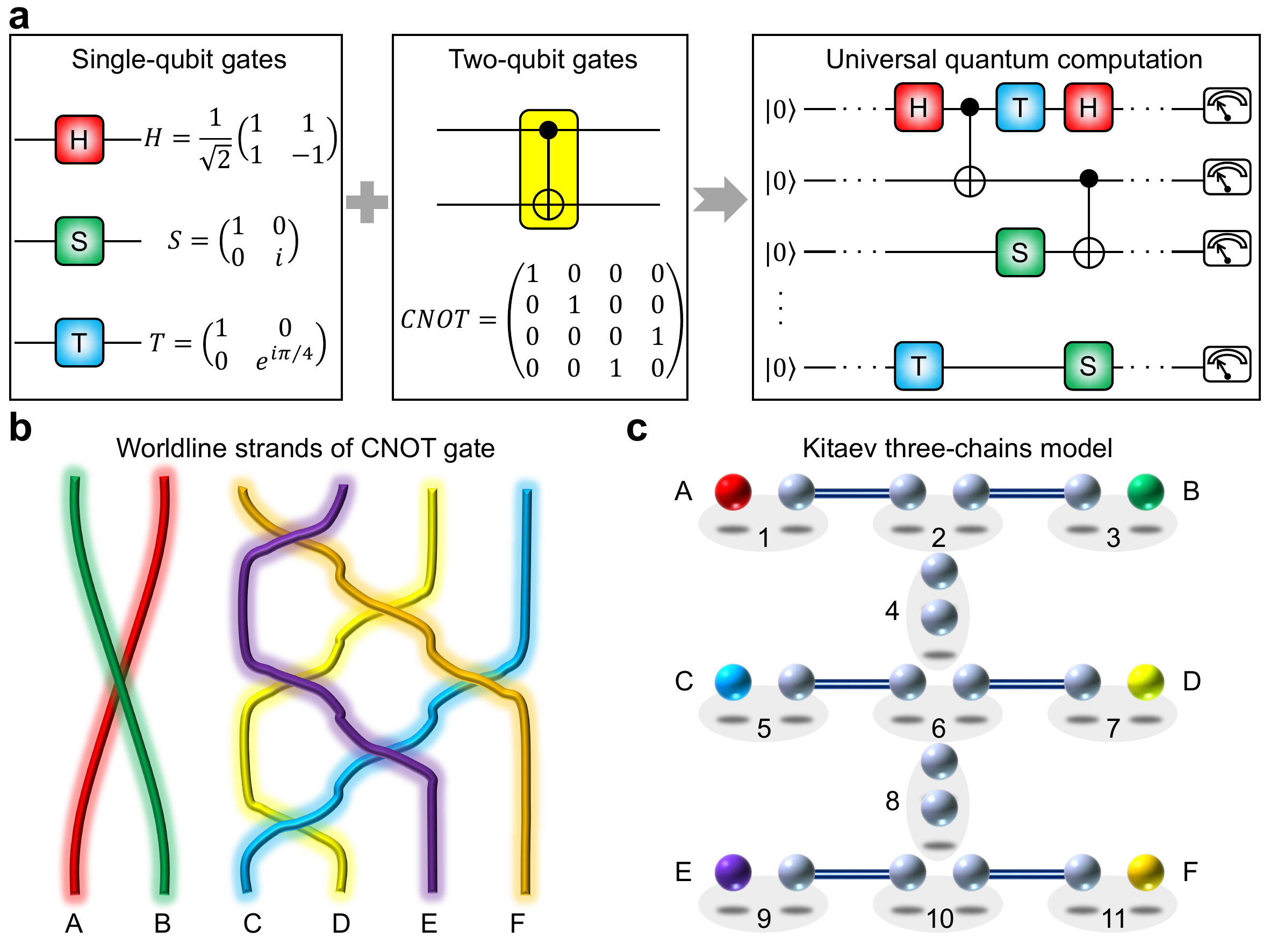}
 	\caption{Theoretical schematic. \textbf{a}. Schematic diagram for achieving universal quantum computation. By combining simple single-qubit gates (Hadamard gate, S gate and T gate) with two-qubit CNOT gates, universal quantum computation can be realized. \textbf{b}.  Worldline strands of the CNOT gate. Through the encoding of two logical qubits using six Majorana zero modes from A to F, the CNOT gate can be implemented via specific braiding operation between them.   \textbf{c}. Kitaev three-chain model. The chain used to implement braiding operations of the CNOT gate consists of eleven fermions with six Majorana zero modes. Each spheres represents a Majorana fermion, and a pair of Majorana fermions (marked by the gray circle) consitudes a Dirac fermion. The ends of three Kitaev chains host Majorana zero modes, labeled as A-F. }
 	\label{fig:1}
 \end{figure}

The theoretical basis for universal quantum computation is well-established, relying on the combination of single-qubit gates (Hadamard, S, and T gates) and two-qubit gates like the CNOT gate, as shown in Fig. \ref{fig:1}\textbf{a}. In this work, we present the experimental realisation of a robust CNOT gate using MZMs simulated on a photonic platform. The CNOT gate is realised by manipulating MZMs through specific braiding operations, with each logical qubit encoded in a pair of Majorana modes, as shown in Fig. \ref{fig:1}\textbf{b}. These MZMs are encoded in a system of three Kitaev chains, each supporting MZMs at their endpoints, as shown in Fig. \ref{fig:1}\textbf{c}. This approach not only demonstrates the feasibility of entangling topological quantum gates but also highlights the robustness of our photonic platform. Notably, the topological protection further enhances the fidelity of the CNOT gate, exceeding 0.992. At the same time, the use of a photonic platform introduces several advantages, including immunity to thermal noise and precise control over quantum states. In our experiment, correlated photon pairs are used to simulate MZMs, allowing for highly controlled braiding operations that mimic the behavior of topological qubits. Our work demonstrates significant progress in the experimental realisation of topological quantum gates and underscores the advantages of combining Majorana fermions with photonic platforms for robust quantum computation.\\

\noindent
\textbf{Results}

To ensure continuous topological protection, we select chain sizes that are sufficiently large to prevent MZMs from ever occupying the same site during the braiding process, as illustrated in Fig.1\textbf{c}.
Each sphere represents a Majorana fermion, while a Dirac fermion is formed by the combination of two adjacent spheres.  
In this context, the fermions are described with the canonical operators $c_{j}$ and  $c_{j}^{\dagger}$, with  $j$ ranges from 1 to 11. The first chain consists of $j$=1, 2, 3; the second chain is composed of $j$=5, 6, 7; and the third chain includes $j$=9, 10, 11. The sites at $j$=4 and $j$=8 serve as connectors between these chains, enabling the braiding process.  The encoded Majorana modes' Hamiltonian within the Kitaev chains can be written as
\begin{equation}
	\begin{split}
		&H_{M_0}=i(\gamma_{1b} \gamma_{2a}+\gamma_{2b} \gamma_{3a}+\gamma_{5b} \gamma_{6a}+\gamma_{6b} \gamma_{7a}+\gamma_{9b} \gamma_{10a}+\\&\gamma_{10b}\gamma_{11a}+\gamma_{4a}\gamma_{4b}+\gamma_{4a}\gamma_{4b}),
		\label{eqn:0}
	\end{split}
\end{equation} 
where $\gamma_{ja}=c_{j}+c_{j}^{\dagger}$ and $\gamma_{jb}=i(c_{j}^{\dagger}-c_{j})$. The Majorana operators $\gamma_{m}$ satisfies the relations $\gamma_{m}^{\dagger}=\gamma_{m}$ and $\gamma_{l} \gamma_{m} +\gamma_{m} \gamma_{l} =2\delta_{lm}$ for $l,m=1a,1b,...11a,11b$. Notably, the operators $\gamma_{1a}$, $\gamma_{3b}$, $\gamma_{5a}$, $\gamma_{7b}$, $\gamma_{9a}$ and $\gamma_{11b}$ do not appear in the initial Hamiltonian $H_{M_0}$, resulting in $[H_{M_0},\gamma_{j}]=0$ for $j=1a, 3b, 5a, 7b, 9a, 11b$. Consequently, these Majorana modes have zero energy and correspond to six endpoint MZMs, labeled from A to F in Fig. \ref{fig:1}\textbf{c}.  

Firstly, to prepare the quantum states of the CNOT operation experimentally, the MZMs systems are transformed to spin system through the Jordan-Wigner transformation \cite{JW1928} thus enabling our photonic simulation. For simplicity, we represent the spin states of the system using eigenvectors $\{|x\rangle,|\bar{x}\rangle\}$, $\{|y\rangle,|\bar{y}\rangle\}$ and $\{|z\rangle,|\bar{z}\rangle\}$ of respective Pauli operators $\sigma^x$, $\sigma^y$ and $\sigma^z$, each having eigenvalues of $\{1,-1\}$.
Hence, the eigenstates of the ground states from the first Kitaev chain to the third Kitaev (which is denoted as $|0_{1}\rangle, |1_{1}\rangle,|0_{2}\rangle, |1_{2}\rangle$ and $|0_{3}\rangle, |1_{3}\rangle$) to Majorana-based spin basis are written by $|0_{1}\rangle=(|x_{1}x_{2}x_{3}\rangle+|\bar{x}_{1}\bar{x}_{2}\bar{x}_{3}\rangle)/\sqrt{2}$, $|1_{1}\rangle=(|x_{1}x_{2}x_{3}\rangle-|\bar{x}_{1}\bar{x}_{2}\bar{x}_{3}\rangle)/\sqrt{2}$, $|0_{2}\rangle=(|x_{5}x_{6}x_{7}\rangle-|\bar{x}_{5}\bar{x}_{6}\bar{x}_{7}\rangle)/\sqrt{2}$, $|1_{2}\rangle=(|x_{5}x_{6}x_{7}\rangle+|\bar{x}_{5}\bar{x}_{6}\bar{x}_{7}\rangle)/\sqrt{2}$, $|0_{3}\rangle=(|x_{9}x_{10}x_{11}\rangle-|\bar{x}_{9}\bar{x}_{10}\bar{x}_{11}\rangle)/\sqrt{2}$ and $|1_{3}\rangle=(|x_{9}x_{10}x_{11}\rangle+|\bar{x}_{9}\bar{x}_{10}\bar{x}_{11}\rangle)/\sqrt{2}$. 
     
\begin{figure*}[htbp]
	\centering
	\includegraphics[width=2\columnwidth]{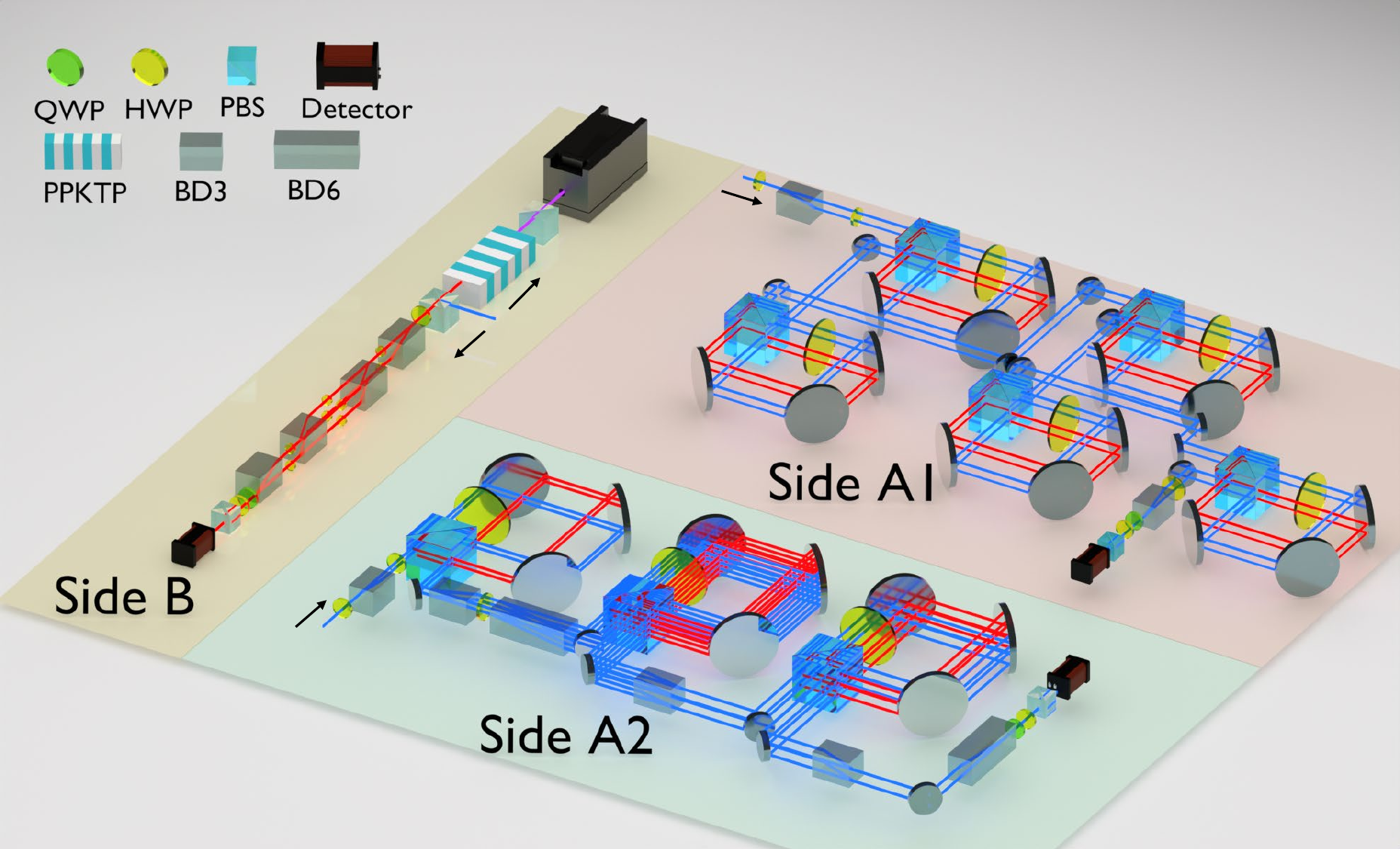}
	\caption{Experimental setup. Correlated photon pairs are generated through the type-\uppercase\expandafter{\romannumeral+2} spontaneous down conversion process. The two photons are then sent to Side A1/A2 and Side B. The two fundamental operations for implementing the CNOT gate: intra-chain braiding and inter-chain braiding, are performed as follows: The collaborative effort of the device in Side A1 and Side B realizes the inter-chain braiding evolution, whereas the device in Side A2 and Side B collaborates to accomplish the intra-chain braiding evolution. Quarter-wave plate (QWP), half-wave plate (HWP), polarization beam splitter (PBS), periodically poled  KTiOPO$_{4}$ (PPKTP), beam displacer (BD).   }
	\label{fig:2}
\end{figure*}
\begin{figure}[htbp]
	\centering
	\includegraphics[width=1\columnwidth]{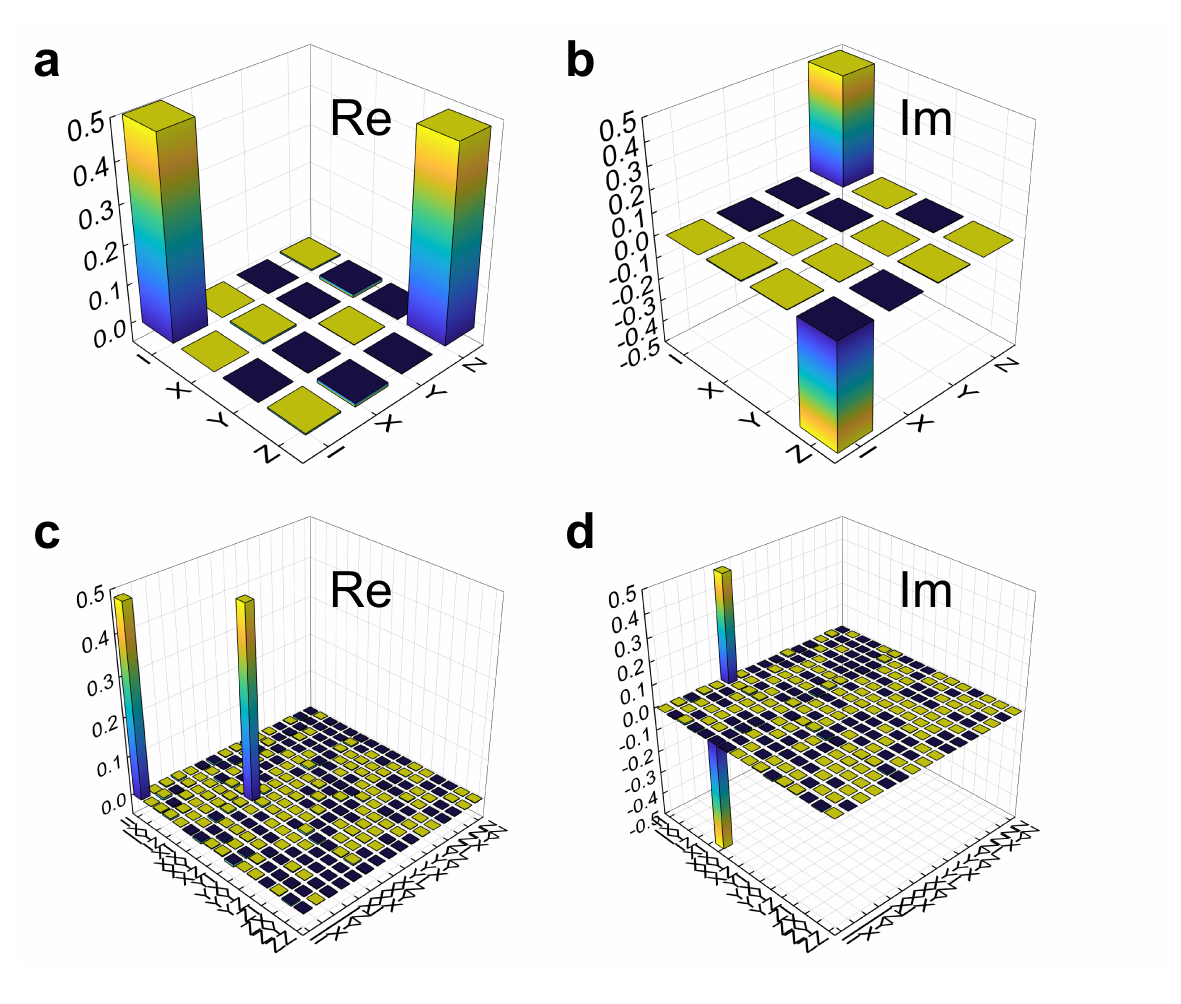}
	\caption{Experimental reconstructed density matrix for intra-chain exchange and inter-chain exchange braiding operation. $\textbf{a}$. Real (Re) and  $\textbf{b}$. imaginary (Im) parts of the process density matrix $\chi_{intra}=(I-iZ)/\sqrt{2}$ of intra-chain exchange braiding operation $\sigma_{1}$. $\textbf{c}$. Real (Re) and  $\textbf{d}$. imaginary (Im) parts of the process density matrix $\chi_{inter}=(II-iXX)/\sqrt{2}$ of inter-chain exchange braiding operation.}
	\label{fig:3}
\end{figure}

In this work, two-photon correlations \cite{PhysRevB.66.045308,PhysRevLett.112.143604} are implemented to encode the quamtum states of CNOT operation. The experimental setup is shown in Fig. \ref{fig:2}. To prepare the two correlated photon pairs, we send a continuous laser pulse with $\lambda=404$ nm to pump a type-\uppercase\expandafter{\romannumeral+2} periodically poled  KTiOPO$_{4}$ (PPKTP), where the induced spontaneous parametric down conversion (SPDC) process \cite{Eisaman2011InvitedRA} generates a pair of photons with the wavelength of 808 nm. One of the photons is sent to side B, and the other photon is sent to side A1/A2, as labelled in Fig. \ref{fig:2}. Two types of beam displacers (BD3 with a displacement of 3 mm and BD6 with a displacement of 6 mm) are employed to spatially separate the photon according to its polarization state, which can be adjusted through the combination of the half-wave plates (HWPs) and quarter-wave plates (QWPs). Cascaded Sagnac interferometers (SIs) in side A1/A2 are utilized to perform the non-dissipative imaginary-time evolution (ITE). The coincidence of the spatial modes between the two photons distributed in the two sides jointly encode a specific qubit during the evolution in the spin basis. For instance, the initial state $|000\rangle$ (ignore subscript) is represented in terms of spin basis as $|000\rangle = (|x_{1}x_{2}x_{3}\rangle+|\bar{x}_{1}\bar{x}_{2}\bar{x}_{3}\rangle)/\sqrt{2} \otimes (|x_{5}x_{6}x_{7}\rangle-|\bar{x}_{5}\bar{x}_{6}\bar{x}_{7}\rangle)/\sqrt{2} \otimes (|x_{9}x_{10}x_{11}\rangle-|\bar{x}_{9}\bar{x}_{10}\bar{x}_{11}\rangle)/\sqrt{2}$. The coincidence of the spatial modes in side A2 and side B will encode each item of this spin state. Here, we denote the spatial modes in side A2 and side B as $\ket{n}_{A2}$ and $\ket{m}_{B}$, respectively, where $n \in \{1,2\}$ and $m \in \{a,b,c,d\}$. Thus, the two-photon correlation modes are expressed as $\ket{nm}=\ket{n}_{A2} \otimes \ket{m}_{B}$, which corresponds to the eight items of $|000\rangle$ expanded in spin basis. The encoding methods of other quantum states are similar to this.  (Specific correspondences can be found in the Supplemental Material.)

 \begin{figure}[!t	]
 	\centering
 	\includegraphics[width=1\columnwidth]{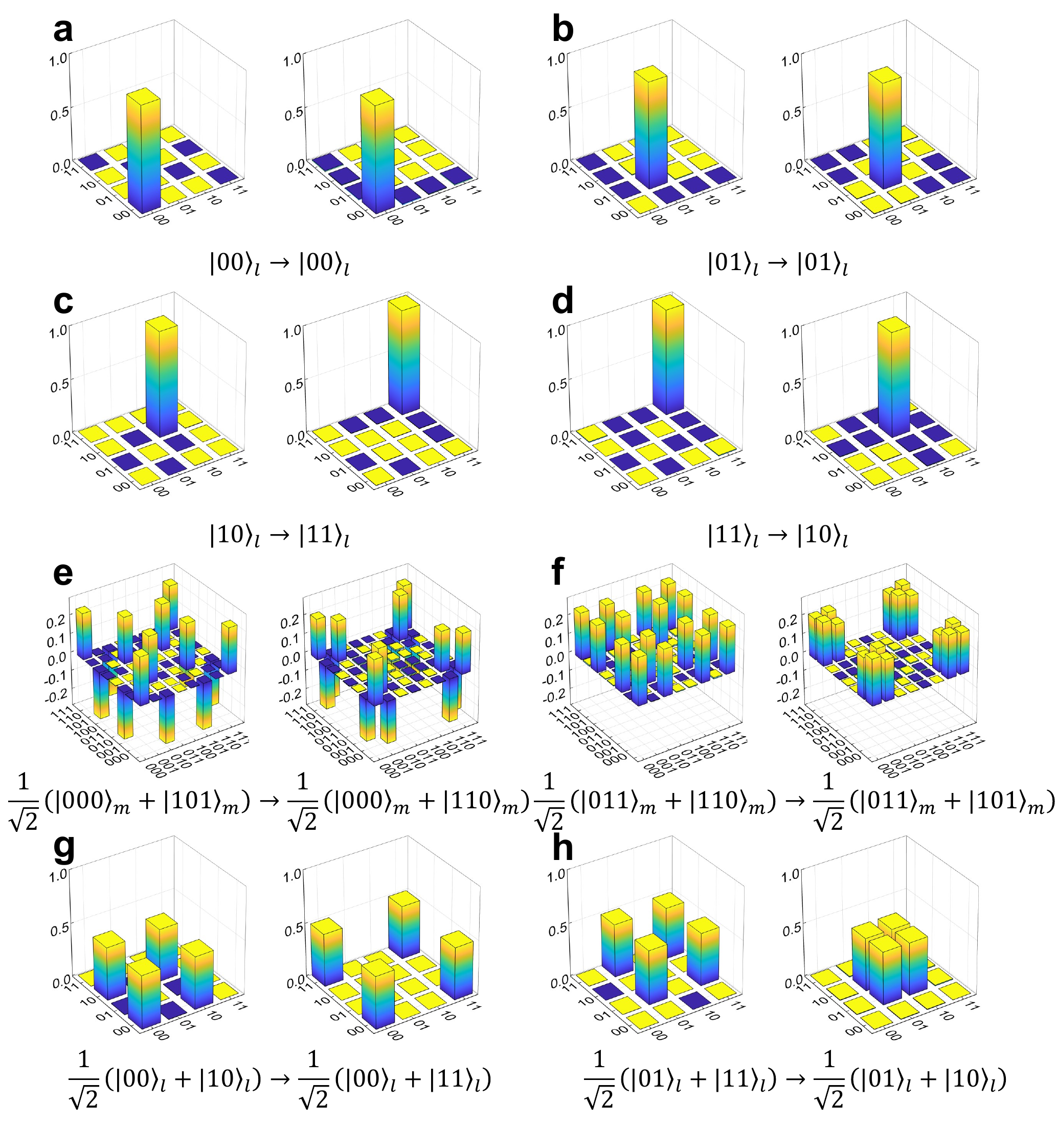}
 	\caption{Experimental results for the input states and the corresponding output states after the action of the CNOT gate. $\textbf{a}$-$\textbf{d}$. Real parts of the input states $|00\rangle_{l}$, $|01\rangle_{l}$, $|10\rangle_{l}$ and $|11\rangle_{l}$ with the corresponding output states $|00\rangle_{l}$, $|01\rangle_{l}$, $|11\rangle_{l}$ and $|10\rangle_{l}$, respectively. $\textbf{e}$-$\textbf{h}$. Experimental results for entanglement generation through the CNOT gate.  The real parts of the density matrix in measurement basis ($\textbf{e}$ and $\textbf{f}$) and logical basis ($\textbf{g}$ and $\textbf{h}$).  The imaginary parts of these density matrix are too small thus not shown in the figure.}
 	\label{fig:4}
 \end{figure}
 
The overall system composed of the three Kitaev chains satisfies the conservation of Fermi parity, where in spin basis, the even parity space composed of $|000\rangle$, $|011\rangle$, $|101\rangle$ and $|110\rangle$, and the odd parity space composed of $|001\rangle$, $|010\rangle$, $|100\rangle$ and $|111\rangle$. The even (odd) space consisting of three Kitaev chains contains four distinct states, which can encode two topological qubits. Thus, here we focus on the even parity and then prepare four input states: $|000\rangle$, $|011\rangle$, $|101\rangle$ and $|110\rangle$, which corresponds to logical states  $|00\rangle_{l}$, $|01\rangle_{l}$, $|10\rangle_{l}$ and $|11\rangle_{l}$. The average initial state fidelity  ($F_{s}=Tr[\sqrt{\sqrt{\rho_{th}}\rho_{exp}\sqrt{\rho_{th}}}]^{2}$) in spin and logical basis is 0.9884 $\pm$ 0.0137 and 0.9976 $\pm$ 0.0139, respectively, which demonstrates the high degree of accuracy in our initial state preparation.

Next, we need to carry out the CNOT gate upon the initial input states. In Majorana-based system, the CNOT operation is implemented via a series of imaginary time evolution operators acting on the initial input state  $\ket{\phi_{in}}$,  which is written as $L_{\sigma^{-1}_{2}}L_{\sigma_{3}}L_{\sigma_{4}}L_{\sigma_{2}}L_{\sigma_{3}}L_{\sigma^{-1}_{2}}L_{\sigma_{1}}\ket{\phi_{in}}$, where $L_{\sigma_{1}}=(\mathbb{1}-i\sigma_{z}) \otimes \mathbb{1}\otimes\mathbb{1}$,$L_{\sigma_{2}}=\mathbb{1}\otimes(\mathbb{1}-i\sigma_{z})\otimes\mathbb{1}$, $L_{\sigma^{-1}_{2}}=  \mathbb{1}\otimes(\mathbb{1}+i\sigma_{z})\otimes\mathbb{1}$, $L_{\sigma^{3}}=  \mathbb{1}\otimes(\mathbb{1} \otimes \mathbb{1}-i\sigma_{x}\otimes\sigma_{x})/\sqrt{2}$ and $L_{\sigma_{4}}=  \mathbb{1}\otimes\mathbb{1}\otimes(\mathbb{1}-i\sigma_{z})$. The standalone identity matrix $\mathbb{1}$ signifies the chain that is not involved in the braiding process.  Furthermore, it can be seen that these operators can actually be classified into two types: Intra-chain exchange and inter-chain exchange.  The intra-chain exchange refers to the exchange between the MZMs within the same Kitaev chain (e.g., A-B, C-D, E-F), while the inter-chain exchange refers to the exchange between the MZMs located in different Kitaev chains (e.g., D-E). The resulting operation of clockwise braiding of intra-chain exchange and inter-chain exchange in logical basis is given by $U_{intra}=\begin{pmatrix} 1&0\\0&i\end{pmatrix}=\mathbb{1}-i\sigma_{z}$ and $U_{inter}=\begin{pmatrix} 1&0&0&-i\\0&1&-i&0\\0&-i&1&0\\-i&0&0&1\end{pmatrix}/\sqrt{2}=(\mathbb{1} \otimes \mathbb{1}-i\sigma_{x}\otimes\sigma_{x})/\sqrt{2}$, respectively. The operation matrix corresponding to the counterclockwise exchange and clockwise exchange of the same MZMs are mutual inverses of each other. Overall, these two basic kinds of operations consists of the aforementioned imaginary time evolution operators, thus achieving the whole braiding process of CNOT gate. Here, we constructed an interferometer comprised of five cascaded Sagnac interferometers on side A1, which, together with the device on side B, simulates the evolution of inter-chain exchange. And the intra-chain exchange is realized through the setup on side A2 together with the device on side B. Several BD3 and BD6 were employed in the  within the cascaded Sagnac interferometers. In order to ensure the effectiveness of these two basic evolutions, we perform quantum process tomography to characterize $U_{intra}$ and $U_{inter}$. A quantum process $\varepsilon$ which is operated on a single-qubit density matrix $\rho$ can be written as $\varepsilon(\rho)= \displaystyle \sum^{4}_{m,n=1}\chi_{mn}\hat{E}_{m}\rho\hat{E}_{n}^\dagger$, where $\hat{E}_{m}$($\hat{E}_{n}$) is the Pauli operators $\left\{I,X,Y,Z\right\}$. For the two-qubit occasion, $\varepsilon(\rho)= \displaystyle \sum^{16}_{m,n=1}\chi_{mn}\hat{E}_{m}\rho\hat{E}_{n}^\dagger$, where $\hat{E}_{m}$($\hat{E}_{n}$) is the tensor product of Pauli operators $\left\{I,X,Y,Z\right\}\otimes\left\{I,X,Y,Z\right\}$. The experimental reconstructed process matrix is shown in Fig. \ref{fig:3}.
  With the definition of the process fidelity $F_{p}=Tr[\sqrt{\sqrt{\chi_{exp}}\chi_{ideal}\sqrt{\chi_{exp}}}]^{2}$, we obtain the process fidelity for $\chi_{intra}=(I-iZ)/\sqrt{2}$ and $\chi_{inter}=(II-iXX)/\sqrt{2}$ is 0.991 $\pm$ 0.014 and 0.961 $\pm$ 0.014, respectively, which demonstrated the high performance of the constructed experimental setup. 
    
   Finally, after realizing the two basic braiding operations consist of the CNOT gate, we perform the quantum state tomography upon the final output states $\ket{\phi_{out}}$ following the application of the CNOT gate. The obtained three-qubit quantum states have 8 basic state vectors in measurement basis from $\ket{000}_{m}$ to $\ket{111}_{m}$, and their density matrix are reconstructed through standard three-qubit quantum state tomography with the tensor products of the measurement basis $\ket{0}_{m},\ket{1}_{m},\ket{+}=(\ket{0}_{m}+\ket{1}_{m})/\sqrt{2}$ and $\ket{R}=(\ket{0}_{m}-i\ket{1}_{m})/\sqrt{2}$. Furthermore, it should be mentioned that, ideally, to experimentally demonstrate the evolution of a CNOT gate, multiple devices resembling side A1/A2 and side B would need to be further cascaded according to the evolution sequence $L_{\sigma^{-1}_{2}}L_{\sigma_{3}}L_{\sigma_{4}}L_{\sigma_{2}}L_{\sigma_{3}}L_{\sigma^{-1}_{2}}L_{\sigma_{1}}\ket{\phi_{in}}$ (See Supplemental Material for the detailed derivation). However, given the current technological limitations, this poses an extremely challenging task to achieve. Therefore, in this work, we adopt an approach where the final state of each evolution step is used as the initial input state for the next evolution step. For example, for the last step evolution ($L_{\sigma^{-1}_{2}}$), we use the final state after evolution $L_{\sigma_{3}}$  which is then sent to the device on side A1 and side B as the initial state. In this case, since the initial state is completely re-prepared, it is considered to have a state fidelity close to 1. However, under this approach, the resulting final state may fail to reflect the influence of previous evolution steps through multiple devices, resulting in an excessively high state fidelity. Therefore, it is necessary to encode the effect of errors to the final state originating from the implementation of each step. In the experiment, imperfect interference of the cascaded interferometer is the primary factor that lead to the reduction the fidelity of the quantum states. The decoherence affects the quantum state through the correlated dephasing channel \cite{Gao_2023}:  $\varepsilon(\rho)=(1-p)\rho+p(Z\otimes Z\otimes Z)\rho(Z\otimes Z\otimes Z)$, where $\rho$ is the density matrix of the input state and $Z$ is the $\sigma_z$ Pauli matrix. By minimizing the discrepancy between the theoretical values and the corresponding experimental data, we obtained the value of $p$ for all the final states in the experiment, and finally calculated the average to yield $p = 0.012$ (See Supplemental Material for the detailed method). After correction, the average state fidelity of the four output states in spin and logical basis is 0.8974 $\pm$ 0.0033 and 0.9901 $\pm$ 0.0036, respectively. The density matrix of the initial states and their corresponding output states following the action of the CNOT gate are shown in Fig. \ref{fig:4}\textbf{a}-\textbf{d}. 
   
One of the most significant and fundamental role of a CNOT gate is to generate entangled states. Here, we entangle the initially separated states to the entangled state with the application of the CNOT gate. Specifically, two typical Bell states $\Phi^{+}=(\ket{00}_{l}+\ket{11}_{l})/\sqrt{2}$ and $\Psi^{+}=(\ket{01}_{l}+\ket{10}_{l})/\sqrt{2}$ are generated by inputting separated states $(\ket{0}_{l}+\ket{1}_{l})\ket{0}_{l}/\sqrt{2}=(\ket{00}_{l}+\ket{10}_{l})/\sqrt{2}$ and $(\ket{0}_{l}+\ket{1}_{l})\ket{1}_{l}/\sqrt{2}=(\ket{01}_{l}+\ket{11}_{l})/\sqrt{2}$. The obtained fidelity of states $\Phi^{+}$ and $\Psi^{+}$ in logical basis with fidelities of 0.9938 $\pm$ 0.0013 and 0.9953 $\pm$ 0.0044, respectively. The obtained high fidelities of the output states indicate the entanglement generation capability of the CNOT gate. The experimental results of the input/output states both in spin and logical basis are shown in Fig. \ref{fig:4}\textbf{e}-\textbf{h}.
  
  \begin{figure}[t]
  	\centering
  	\includegraphics[width=1\columnwidth]{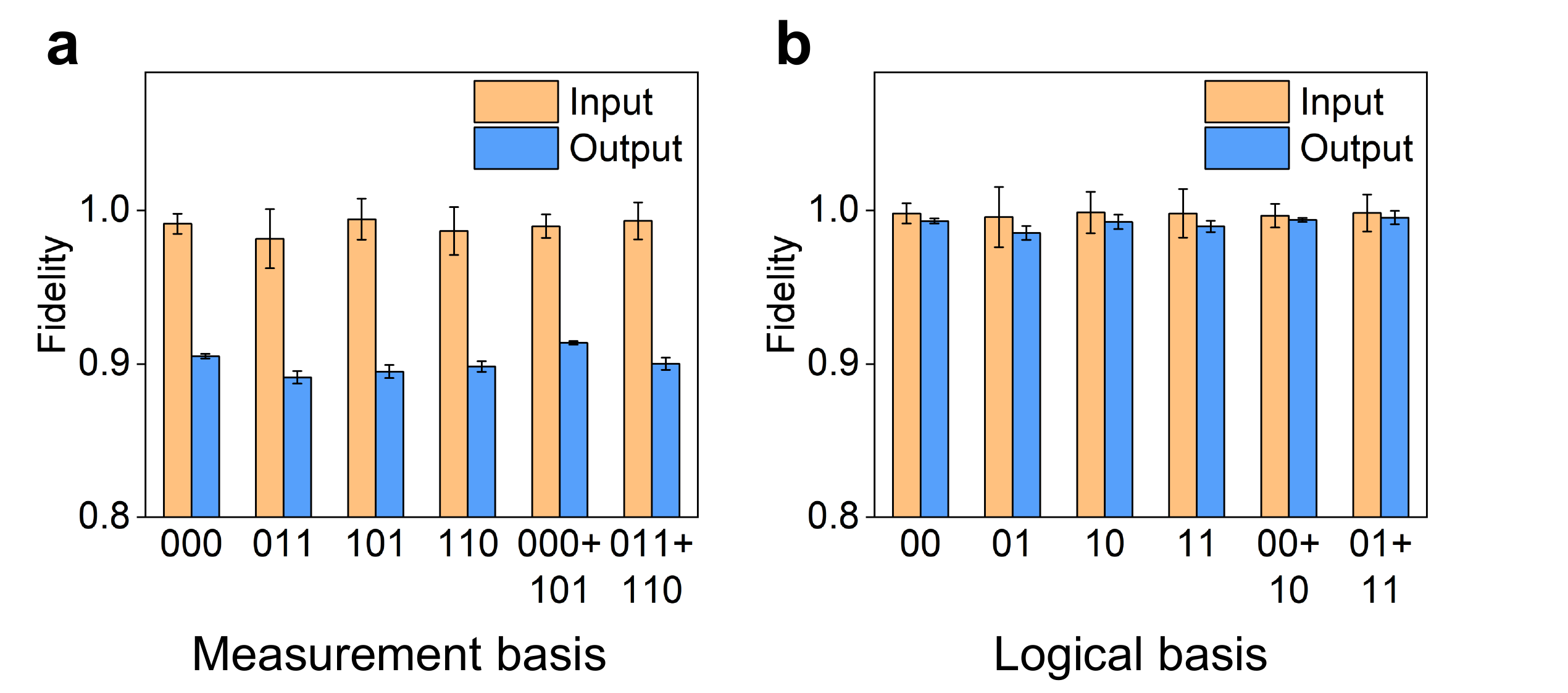}
  	\caption{Comparison of the input/output state fidelities under different basis. \textbf{a}. Input states from $|001\rangle_{m}$  to $(|011\rangle_{m}+|110\rangle_{m})/\sqrt{2}$ with their corresponding output states after the action of the CNOT gate in measurement basis. \textbf{b}. Input states from $|00\rangle_{l}$ to $(|01\rangle_{l}+|11\rangle_{l})/\sqrt{2}$ with their corresponding output states in logical basis.  }
  	\label{fig:5}
  \end{figure}  
    
We subsequently encode two-qubit logical states using three-qubit spin states in a system of three Kitaev chains. This approach  borrows the core idea from fault-tolerant quantum computing schemes, which advocate the utilization of a larger number of physical qubits to encode a comparatively small quantity of logical qubits. This serves to potentially mitigate errors and enable a higher probability of correct output in quantum computations. Here, we demonstrate the quantum state fidelities for all the input/output states both in measurement/logical basis. Input states from  $|001\rangle_{m}$ ($|00\rangle_{l}$) to $(|011\rangle_{m}+|110\rangle_{m})/\sqrt{2}$ $(|01\rangle_{l}+|11\rangle_{l})/\sqrt{2})$ in measurement (logical) basis and their corresponding output states are shown in Fig. \ref{fig:5}\textbf{a} and \textbf{b}, respectively. The average fidelity of the input (output) states in measurement basis is 0.9894 $\pm$ 0.0124 (0.9006 $\pm$ 0.0031). Meanwhile, the average fidelity of the input (output) states in logical basis reaches 0.9976 $\pm$ 0.0126 (0.9916 $\pm$ 0.0034),  representing a further improvement compared to that in the measurement basis. The obtained high fidelities validates the feasibility and unique advantages of our encoding method.

\begin{figure}[t!]
    \centering
    \includegraphics[width=\linewidth]{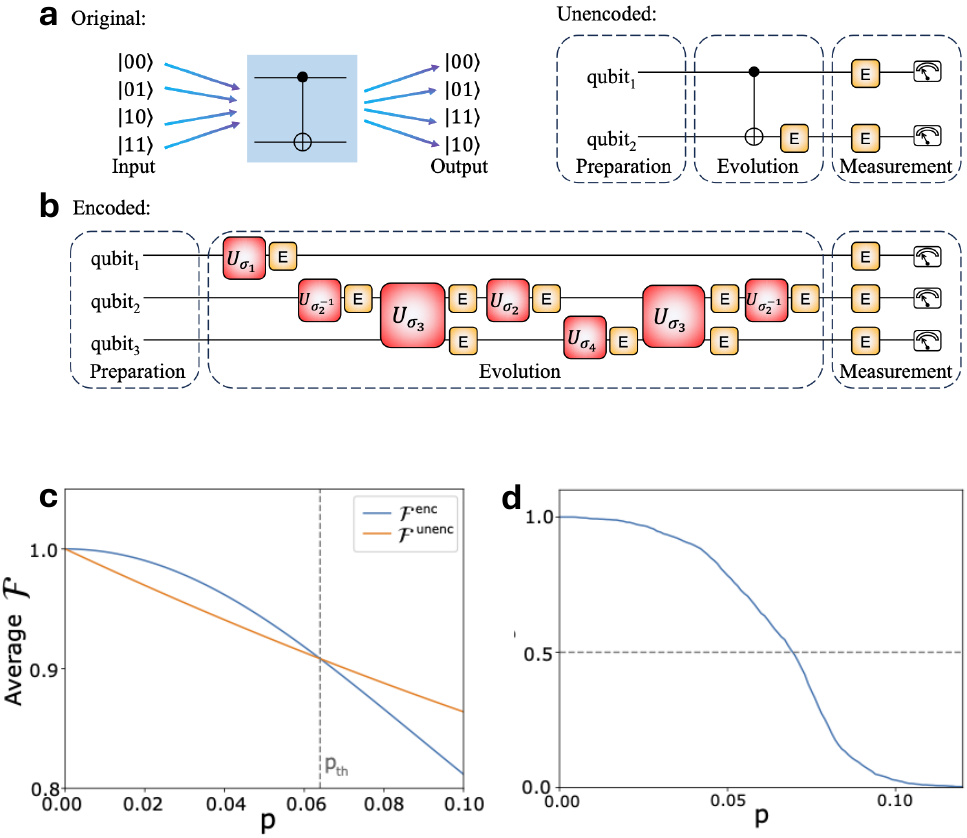}
    \caption{Schematics of {\bf a} the unencoded CNOT gate and {\bf b} the CNOT gate encoded in three-qubit circuit implementing in the presence of local errors, $E$. {\bf c}. Average fidelity $\mathcal{F}$ vs. error probability $p$ for encoded and unencoded CNOT gates over 2000 Haar-random states. Encoding yields higher fidelity for $p < p_{\text{th}}$ where $p_{\text{th}}\approx 0.065$. {\bf d}. Probability $P_{\rm avg}$ that the encoded fidelity exceeds on average the unencoded fidelity across 2000 input states vs. $p$. A clear advantage emerges below $p \approx 0.07$.}
    \label{fig:m1}
\end{figure}

We now investigate further the resilience of the CNOT gate encoded in MZM operations against experimentally induced decoherence. While topological encoding provides intrinsic protection against local perturbations, such as coherent errors or Hamiltonian fluctuations, this protection does not necessarily apply to the errors that are naturally present in a given physical implementation. This discrepancy arises due to the Jordan-Wigner transformation used in our encoding where local fermionic errors can be mapped to non-local spin errors, altering their effect on the system. 

In earlier work~\cite{Xu2018}, we demonstrated that topological encoding using MZMs is inherently fault-tolerant under local fermionic perturbations. This fault tolerance is a consequence of non-Abelian statistics: local operators cannot change the total fermionic parity or induce transitions between distinct topological sectors. However, these fermionic-local errors do not directly correspond to the local spin errors typically present in any experimental implementation.

Here, we show that the MZM encoding protects against a different and more experimentally relevant class of errors, namely those originating from environmental decoherence and imperfections in the photonic hardware. Notably, we find that the topologically encoded CNOT gate in our photonic system exhibits enhanced resilience to such errors, provided the error probability remains below a well-defined threshold. This form of resilience has no direct analogue in conventional topological protection, and it highlights the practical utility of our approach under realistic noise conditions.

To assess this robustness, we compare the performance of the topologically encoded CNOT circuit (implemented using three physical qubits) with that of a conventional, unencoded CNOT gate acting on two qubits. In both cases, we introduce uncorrelated, local unitary errors, each acting on participating qubits with probability $p$. The noisy operation is modelled as
\begin{equation}
\epsilon(O, U_E, p) = (1 - p)O + p U_E O U_E^{\dagger},
\end{equation}
where $U_E$ represents the error channel, taken here to be $Z$, corresponding to decoherence. Note that instead of the correlated dephasing errors we previously employed when investigating the photonic platform, we now consider uncorrelated errors, allowing for a more general and platform-independent assessment of the encoding’s fault tolerance.

Fig.~\ref{fig:m1}{\bf a} and {\bf b} illustrate the unencoded and the encoded circuits of the CNOT gate, respectively. For each input state $\ket{\psi_{\text{in}}}$, we compute the output density matrix $\rho_{\text{out}}^{\text{enc}}$ for the encoded circuit and $\rho_{\text{out}}^{\text{un}}$ for the unencoded one. The fidelity relative to the ideal output $\rho_{\text{out}} = \text{CNOT}\ket{\psi_{\text{in}}}\bra{\psi_{\text{in}}}\text{CNOT}^\dagger$ is given by
\begin{equation}
\mathcal{F}^{\text{enc/un}} = \sqrt{\text{Tr}(\rho_{\text{out}}\rho_{\text{out}}^{\text{enc/un}})}.
\end{equation}
To obtain a statistically meaningful comparison, we generate 2000 random input states by applying Haar-random unitaries to a fixed reference state $\ket{\psi_0}$. For each input, we evaluate the fidelity under both encoded and unencoded circuits subject to decoherence errors. The average fidelity, $\mathcal{F}$, for each circuit, as a function of $p$, is then computed.

As shown in Figure~\ref{fig:m1}{\bf c}, the encoded CNOT circuit outperforms the unencoded one when the error rate is below a threshold $p_{\text{th}} \approx 0.065$. To further quantify this improvement, we compute the probability $P_{\rm avg}$ that the encoded fidelity exceeds the unencoded fidelity over the 2000 input states. Figure~\ref{fig:m1}\textbf{d} shows that $P_{\rm avg} > 0.5$ for $p < 0.07$, indicating that, in this range, the encoded scheme is more likely to deliver higher-fidelity outputs. Importantly, the experimentally estimated error rate for our setup is $p = 0.012$, which lies well within this regime of enhanced resilience.

This clear threshold behaviour highlights a form of passive protection provided by the topological encoding, extending beyond conventional notions of topological fault tolerance. Even in this minimal implementation, the encoding significantly suppresses the impact of decoherence. While the experimental gate operations are already highly accurate, this additional resilience brings us closer to the thresholds required for scalable quantum computing.

\noindent
\textbf{Conclusions}

We have successfully demonstrated the experimental realisation of a robust CNOT gate using a Majorana-based photonic platform. By encoding logical qubits in Majorana zero modes and employing both intra-chain and inter-chain braiding operations, we achieved high-fidelity quantum state preparation and gate operation. Our results demonstrate an inherent robustness of topological quantum gates against induced noise and decoherence that naturally occur in the photonic implementation. This is evidenced by the consistently high fidelity of output logical states, including maximally entangled Bell states.
This work provides an essential proof of concept for topological quantum computing, illustrating the benefits of simulating Majorana fermions with photonic platforms. The experimentally obtained high fidelity observed in both spin and logical bases indicates that the system is indeed well-suited for scaling to more complex quantum operations. Additionally, the topological encoding used in our experiment offers promising error mitigation strategies, further supporting the practicality of Majorana-based quantum computation. Future work may focus on scaling this system to multi-qubit operations and exploring the integration of additional quantum gates to implement even more complex quantum information processing tasks. \\

\noindent
\textbf{Acknowledgements}\\
This work was supported by the Innovation Program for Quantum Science and Technology (Grants No. 2021ZD0301200 and No. 2021ZD0301400), the National Natural Science Foundation of China (Grants No. 12374336, No. 92365205, No. W2411001, No. 62475249), USTC Major Frontier Project (LS2030000002) and USTC Research Funds of the Double First-Class Initiative (Grant No. YD2030002024).\\

\noindent
\textbf{Author contributions}\\
Y.-J.H. proposed this project. K.S. designed the experiment. J.-K.L. carried out the 
experiment and analysed the experimental results assisted by K.S., Z.-Y.H. and J.-H.L.. J.K.P contributed to the theoretical analysis assisted by L.B.. J.-K.L. and J.K.P wrote the manuscript.  J.-S.X., Y.-J.H., C.-F.L. and G.-C.G. supervised the project. 
All the authors discussed the experimental procedures and results.\\
\bibliography{reference.bib}

\end{document}